# REVERSIBLE PROGRAMMABLE LOGIC ARRAY (RPLA) USING FEYNMAN & MUX GATES FOR LOW POWER INDUSTRIAL APPLICATION

Pradeep Singhla, Hindu College of Engineering, Sonipat, Haryana, India
pardeep51355@gmail.com
Naveen Kumar Malik, Hindu College of Engineering, Sonipat, Haryana, India
naveenmalik4u@gmail.com

## ABSTRACT

*This paper present the research work directed towards the design of reversible programmable logic array using very high speed integrated circuit hardware description language (VHDL). Reversible logic circuits have significant importance in bioinformatics, optical information processing, CMOS design etc. In this paper the authors propose the design of new RPLA using Feynman & MUX gate. VHDL based codes of reversible gates with simulating results are shown .This proposed RPLA may be further used to design any reversible logic function or Boolean function (Adder, Subtractor etc.) which dissipate very low or ideally no heat.*

## INTRODUCTION

Today's Electronic systems becomes important part of the human's life and engineers wants that the system should have high performance, speed and dissipate very low or ideally no heat. Still the power dissipation is one of the greatest limitation factors in these electronic systems. The digital systems designed by the conventional approach or irreversible design approach consume or dissipate KTln2 of energy on every bit computation *(R. Lndauer, 1961)*. Where K is a Boltzmann's constant equals to $1.3807 \times 10^{-23} \, JK^{-1}$ and T is the temperature at which the computation is performed..The dissipated energy directly correlated to the number of lost bits *(C. H. Bennet, 1973)*. Resultantly, a new design approach has been comes in the hardware designing field for reducing the power dissipation known reversible logic structure possesd the property of one to one mapping between the inputs and output state *(H. R. Bhagyalakshmi, et al., 2010)*.

Reversible logics have very useful applications in everyday life like Laptop computers, wearable computers, spacecraft, smart cards etc *(Michael P. Frank, 2005)*. So, in a new paradigm in computer design, reversible logics play a very important role over irreversible logics. The non-reversible PLA structure may be used to realize the reversible functions *(Ahsan Raja, et al., 2006)*. Programming Logic Devices (PLDs) or Arrays (PLAs) are used in industrial applications to synthesis cost effective solutions to industrial designs *(Hamid R. Arabnia, et al., 2006)*. The PLA designed using reversible gates is called RPLA. The authors seeing the benefits of programmable logic array design approach in industrial application designed new RPLA using reversible gates i.e. Feynman gate & Mux gate. This device can be changed at any time to perform any number of reversible Boolean functions. The proposed new architecture of RPLA has three inputs which can perform $2^8$ logical functions.

The organization of paper as follows: the overview of reversible logic gate is explained first then the various reversible gates used in design are explained. In next the proposed architecture of RPLA is shown. Further, the design of 3 input RPLA with VHDL codes of Feynman & MUX gate with RTL view of Reversible AND plane & Reversible OR plane discussed and shown. The experimental results of Reversible AND plane & Reversible OR plane are shown. Finally Sum up the paper with conclusion and explores the future scopes and enlist the references.

## OVERVIEW OF REVERSIBLE LOGIC

---

The idea of reversible computing comes from the thermodynamics which taught us the benefits of the reversible process over irreversible process. So, a computation is called reversible if its inputs can always be retrieved from its outputs *(H. R. Bhagyalakshmi, et al., 2010)*. But reversible means not only logically reversibility in the circuit but also physical reversibility must be there. Logical reversibility implies that the number of inputs must equal to the outputs *(G. De Mey, et al., 2008)*.

In other words, these circuits can generate unique output vector from each input vector and vice-versa. So, an N×N reversible gate can be represented as *(Ahsan Raja, et al., 2006)*.

$$I_v = (I_1, I_2, I_3, \dots\dots\dots\dots, I_N)$$
$$O_v = (O_1, O_2, O_3, \dots\dots\dots\dots, O_N)$$

Where $I_v$ = input vectors
$O_v$ = output vectors

When a device can actually run backwards then it is called physically reversible and the second law of thermodynamics guarantees that it dissipates no heat.

In this paper, we implement the RPLA by taking the case of logically reversibility. So, we only consider the logical reversibility part because this is the elementary idea which gave rise to a research area about reversible computing *(G. De Mey, et al., 2008)*.

For logical reversibility in the digital logics there are two conditions as follows *(H. R. Bhagyalakshmi, et al., 2010)*.

- Fan-Out is not permitted
- Feedback is not permitted

Some of the important and main measure in designing of the reversible logic circuits are QC (Quantum cost) *(H. R. Bhagyalakshmi, et al., 2010)*: The number of reversible gates (1×1 or 2×2) to realize the circuit is known as quantum cost, (CI) Constant inputs *(H. R. Bhagyalakshmi, et al., 2010)*: The number of inputs that are kept constant (0 or 1) for synthesis the given functions, GO (Garbage outputs) *(H. R. Bhagyalakshmi, et al., 2010)*: The number of outputs that are not primary is known as Garbagr outputs

**Reversible Gates:**

There is several reversible logic gates have been proposed in the last years such as: Feynman gate *(H. R. Bhagyalakshmi, et al., 2010)*, Fredkin gate *(Ahsan Raja, et al., 2006)*, Peres gate *(T. Toffoli, 1980)*, Toffolli gate *(T. Toffoli, 1980)*, new gate *(H. R. Bhagyalakshmi, et al., 2010)*, Mux gate *(Sajib Kumar Mitra, et al., 2011)* etc. Here we are reviewing the three gates i. e. Feynman, Fredkin and MUX gate because these gates are used in the designing of the RPLA.

***Fredkin Gate*** *(Ahsan Raja, et al., 2006)***:** Fredkin Gate is a 3×3 conservative reversible gate. It is called 3×3 gate because it has three inputs and three outputs. The inputs (A, B, C) associates with its outputs outputs (P, Q, R).Its quantum cost is 5.

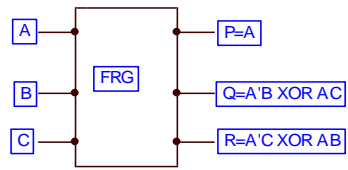

***Figure 1: Fredkin Gate***

***MUX Gate*** *(Sajib Kumar Mitra, et al., 2011)***:** *Figure 2* shows the pictorial representation of 3×3 reversible gate called MUX (MG) Gate. It is a conservative gate havig three inputs (A, B, C) and three outputs (P, Q, R). The outputs are defined by P=A, Q=A XOR B XOR C and R= A'C XOR AB. The hamming weight of its input vector is same as the hamming weight of its output vector and its Quantum cost is 4.

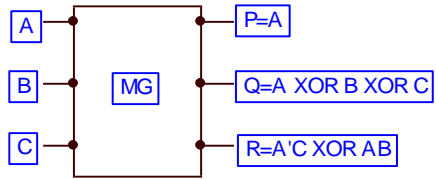

***Figure 2:  3×3 MUX Gate***

***MUX Gate as AND Gate and OR Gate***: The structure of MUX Gate as AND Gate and OR Gate is shown in the *Figure 2.1* and *Figure 2.2*. If we provide '0' at third input C then the output R will provide the AND combination of first & second input and if we If we provide '1' at second input B then the output Q will provide the OR combination of the first & third input.

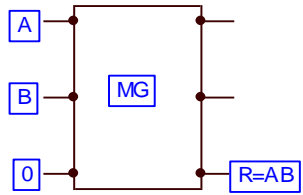
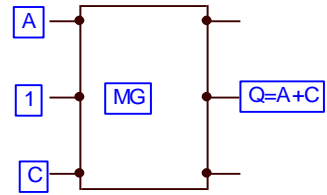

***Figure 2.1: MUX gate as AND gate***            ***Figure 2.2: MUX gate as OR gate***

***Feynman Gate*** *(H. R. Bhagyalakshmi, et al., 2010)***:** *Figure 3* shows the 2×2 reversible gate called Feynman gate. Feynman gate is also recognized as controlled- not gate (CNOT). It has two inputs (A, B) and two outputs (P, Q). The outputs are defined by P=A, Q=A XOR B .This gate can be used to copy a signal. Since fan-out is not allowed in reversible logic circuits, the Feynman gate is used as the fan-out gate to copy a signal. Quantum cost of a Feynman gate is 1.

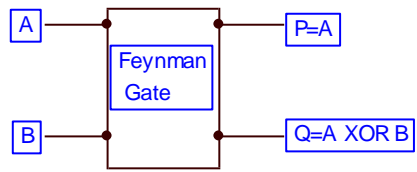

***Figure 3:  2×2 Feynman Gate Structure***

***Feynman Gate as Data Copier & as NOT Gate*** *(H. R. Bhagyalakshmi, et al., 2010)***:** The structure of Feynman gate as Data Copier & as NOT gate is shown in the *Figure 3.1* and *Figure 3.2* respectively. If we provide '0' at second input B then the output Q will provide the copy of first input and if we If we provide '1' at second input B then the output Q will provide the complement of the first input.

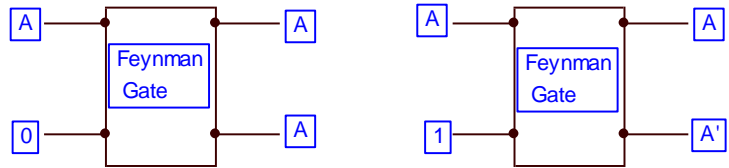

**Figure 3.1: Feynman Gate as Data Copier**     **Figure 3.2: Feynman Gate as NOT Gate**

## PROPOSED ARCHITECTURE OF NEW RPLA:

The RPLA consist of AND plane with buffers/inverters, each of which can be programmed to generate some product terms of input variable combinations with OR plane to realize the output functions. The block representation of RPLA is shown in *Figure 4*.

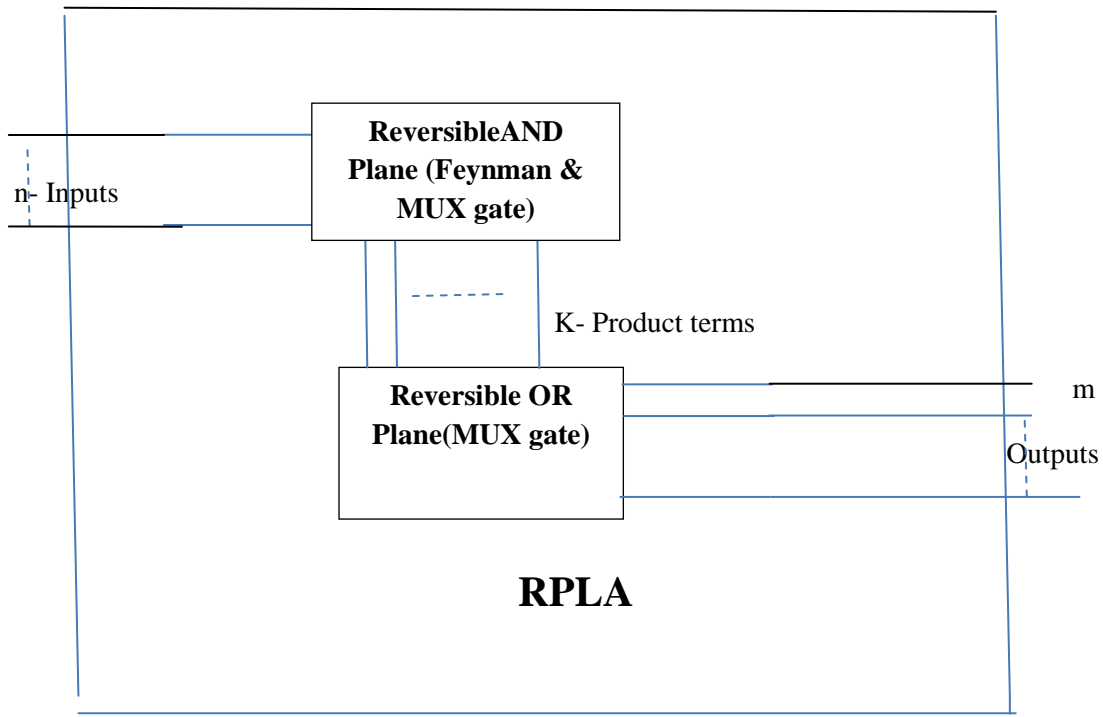

**Figure 4: Schamatic Block Diagram of New RPLA**

In this architecture, we used n numbers of inputs and getting m numbers of outputs as in conventional PLA. In the complete structure of the RPLA there are two planes i.e. reversible AND plane & Reversible OR plane.

## METHODOLOGY

The existing methodology which was proposed by (*Himanshu Thapliyal, et al., 2006*) which uses the two gates for the design i.e. Feynman & Fredkin gate (*Himanshu Thapliyal, et al., 2006*). We propose new methodology for design of RPLA in this paper by using Feynman & MUX gate. The new methodology uses the new gate i.e. MUX gate *(Sajib Kumar Mitra, et al., 2011* having lower quantum cost as compare to the fredkin gate, so by replacing the fredkin gate by the MUX gate the total quantum cost of RPLA decreases. For the designing purpose we use the VHDL and simulate on the xilinx 8.2i.

***VHDL Description:***

This is the language Designed by IBM, Texas Instruments, and Intermetrics as part of the DoD funded VHSIC program and Standardized by the IEEE in 1987: IEEE 1076-1987. Its enhanced version of the defined in 1993: IEEE 1076-1993 *(J. Bhasker)*. The VHDL stands for the very high speed integrated circuit hardware description language having three kinds of modeling i.e. Behavioral Modeling, Structural modeling and Dataflow modeling. For implementing any of the design we go through the following steps *(J. Bhasker)* as shown in Figure 5.

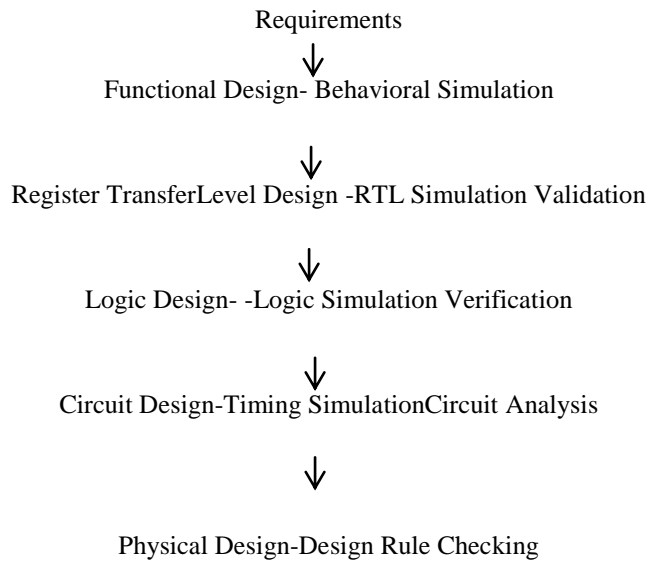

Requirements

↓

Functional Design- Behavioral Simulation

↓

Register TransferLevel Design -RTL Simulation Validation

↓

Logic Design- -Logic Simulation Verification

↓

Circuit Design-Timing SimulationCircuit Analysis

↓

Physical Design-Design Rule Checking

***Figure 5: Basic Flow Steps of a simple Digital design through VHDL***

The proposed design of RPLA uses MUX gate & Feynman gate. The VHDL codes of MUX gate & Feynman gate design in Behavioral modeling are as follows:

***VHDL Codes for MUX gate***
Library IEEE;
Use ieee.std_logic_1164.all;
Use ieee.numeric_std.all;
Entity MG3 is
Port (IN1: in STD_LOGIC;
IN2: in STD_LOGIC;
IN3: in STD_LOGIC;
OUT1: out STD_LOGIC;
OUT2: out STD_LOGIC;
OUT3: out STD_LOGIC);
End MG3;
Architecture Behavioral of MG3 is
Begin
OUT1<= IN1;
OUT2<=IN1 xor IN2 xor IN3;

OUT3<= (((NOT IN1) and IN3) xor (IN1 and IN2));
End behavioral;

***VHDL Codes for Feynman gate***

library ieee;
use ieee.std_logic_1164.all;
use ieee.numeric_std.all;
entity FY2 is
port(IN1:in STD_LOGIC;

IN2:in STD_LOGIC;
OUT1:out STD_LOGIC;
OUT2:out STD_LOGIC);
end FY2;
architecture behavior of FY2 is
begin
OUT1<=IN1;
OUT2<=IN1 xor IN2;
End behavior;

In our research, we define the individual behavior of every gate and then by structural modeling we connect all the output of the previous stage to the next stage by port mapping, assigning them the signal then from the VHDL code the RTL view if AND plane and RTL view of OR plane are generated .These are shown as *Figure 6* and *Figure 7* respectively.

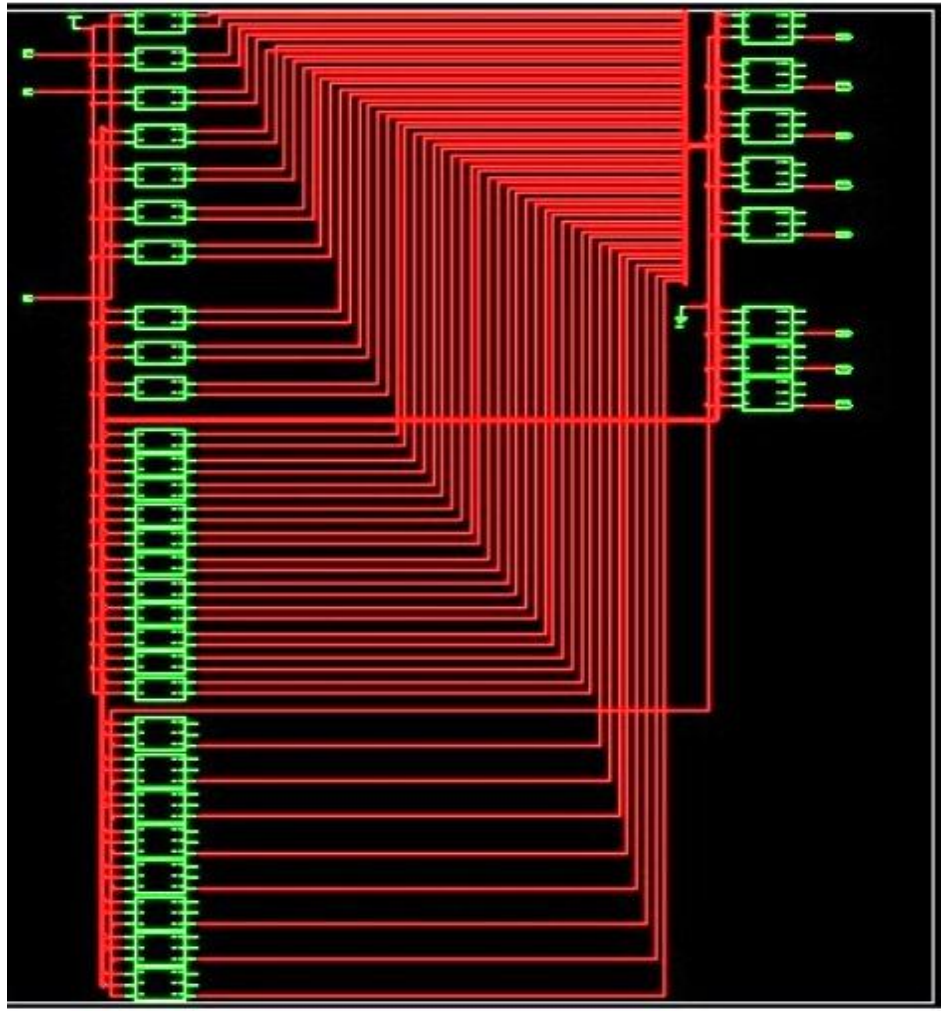

*Figure 6: RTL Schamatic of AND Plane*

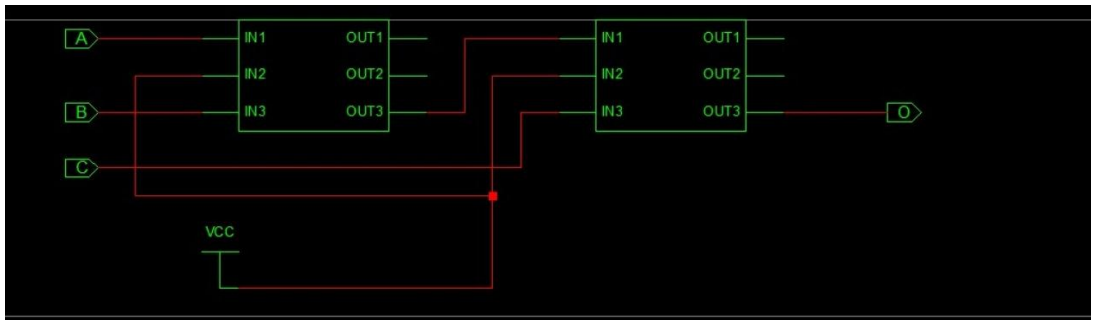

*Figure 7: RTL Schamatic of Reversible OR Plane*

## EXPERIMENTAL RESULT

The proposed new RPLA resulted by VHDL and simulated on Xilinx ISE8.2i.

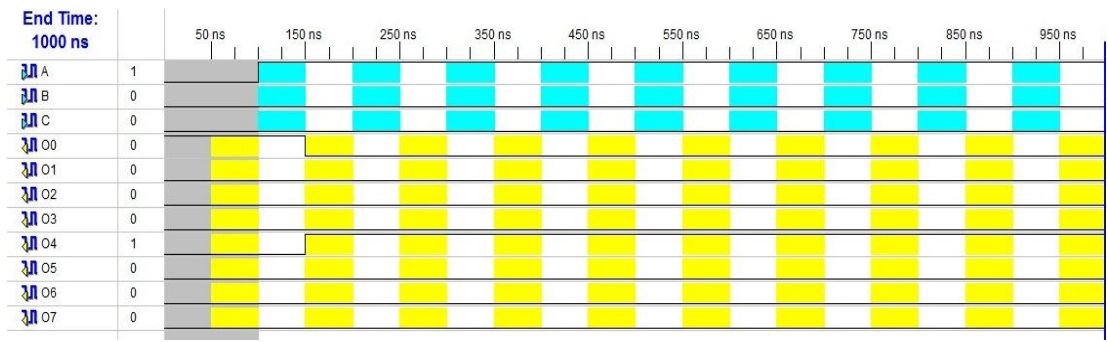

*Figure 8: Simulated result of reversible AND palne*

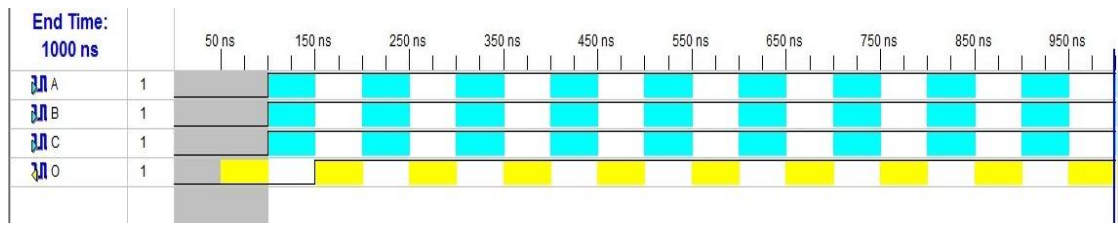

*Figure 9: Simulated result of reversible OR plane*

## CONCLUSION

In this paper we emphasis on an efficient approach to design low power digital system for industrial applications using RPLA. In this new RPLA design by Feynman & Mux gate is proposed..As we already discussed the quantum cost of the MUX gate is lower than the Fredkin gate and Mux gate can perform all operation that can be performed by fredkin. So the proposed RPLA by MUX gate is more cost - efficient than the existing one. In this paper we also represent the simulated results of the proposed RPLA which are completely varified and correct in all sense. So, to reduce the energy consumption, reversible PLA is one of the important logic arrays in digital system design. This work is expected to provide a new approach for design of low power reconfigurable computing for industrial applications.